\documentclass[fleqn,usenatbib]{mnras}

\usepackage{newtxtext,newtxmath}
\newcommand{\source}{PKS 2155-304}

\newcommand{\fermi}{{\it Fermi}-LAT}
\newcommand{\gray}{$\gamma$-ray}
\newcommand{\grays}{$\gamma$-rays}

\usepackage[T1]{fontenc}

\DeclareRobustCommand{\VAN}[3]{#2}
\let\VANthebibliography\thebibliography
\def\thebibliography{\DeclareRobustCommand{\VAN}[3]{##3}\VANthebibliography}

\usepackage{xcolor}    
\usepackage[normalem]{ulem}  

\usepackage{graphicx}	
\usepackage{amsmath}	

\title[Broadband emission from PKS 2155-304]{A comprehensive view of PKS 2155-304 from 2008 to 2023 through a multi-epoch modeling of its spectral energy distributions}

\author[G. Harutyunyan et al.]{
G. Harutyunyan$^{1}$, 
N. Sahakyan$^{1}$, \thanks{E-mail: narek.sahakyan@icranet.org }
D. B\'egu\'e$^{2}$,
M. Khachatryan$^{1}$\\
$^{1}$ICRANet-Armenia, Marshall Baghramian Avenue 24a, Yerevan 0019, Armenia\\
$^{2}$Bar Ilan University, Ramat Gan, Israel
}

\date{Accepted XXX. Received YYY; in original form ZZZ}

\pubyear{\the\year{}}

\begin{document}
\label{firstpage}
\pagerange{\pageref{firstpage}--\pageref{lastpage}}
\maketitle

\begin{abstract}
We present a detailed investigation of the temporal and spectral evolution of
the emission from the blazar PKS 2155-304, a high-synchrotron-peaked blazar. Using $\gamma$-ray, X-ray, optical/UV,
and infrared data assembled from the Markarian Multiwavelength Data Center, we constructed multi-band light curves and temporally resolved
spectral energy distributions (SEDs) of PKS 2155-304 to probe the origin of
its emission. The light curves show significant variability, with fractional
variability peaking at $\sim$0.7–0.8 in soft-to-medium X-rays, $\sim$0.35–0.55
in the optical/UV, and $\sim$0.65 at high-energy $\gamma$-rays-consistent
with expectations for high-synchrotron-peaked blazars. Segmenting the $\gamma$-ray light curve with Bayesian blocks, we defined 253 time-resolved epochs with adequate multi-band coverage
and categorized them into quiescent states (QS), multiwavelength flares (MWF),
$\gamma$-ray flares ($\gamma$F), X-ray flares (XF), and optical/UV flares (OUF).
Each SED is modeled within a one-zone synchrotron self-Compton (SSC) framework
that self-consistently evolves particle injection and cooling; a neural-network surrogate is used to accelerate parameter inference.
Kolmogorov–Smirnov tests reveal state-dependent parameter variations relative
to QS: (i) during MWF, the magnetic field $B$, electron luminosity $L_{\rm e}$, maximum electron Lorentz factor $\gamma_{\rm max}$, and Doppler factor $\delta$ differ significantly; (ii) during $\gamma$F, a harder electron index $p$ is estimated; (iii) XF shows higher $B$ and $\gamma_{\rm max}$ with a more compact emitting region; and (IV) during OUF, changes in $B$, $L_{\rm e}$, $\gamma_{\rm max}$, $\delta$, and $p$ are found while the emitting-zone size remains approximately constant. The jet power is electron-dominated (magnetic-to-electron power ratio $\eta_{\rm B}\simeq0.09$–$0.17$), with $\eta_{\rm B}$ rising during XF. These results suggest that variations in acceleration efficiency and magnetization drive band-dependent flaring in PKS 2155-304.
\end{abstract}

\begin{keywords}
quasars: individual: PKS 2155-304; galaxies: jets; radiation mechanisms: non-thermal;   gamma-rays: galaxies; X-rays: galaxies
\end{keywords}



\section{Introduction}
Blazars are a subclass of radio-loud active galactic nuclei in which a relativistic jet is viewed at a small
angle to the line of sight of the observer \citep{1995PASP..107..803U}. Because of the small viewing angle
and relativistic motion of the jet, relativistic beaming amplifies the jet emission, producing some of the
characteristic blazar features, including: extreme apparent luminosities, rapid broadband variability,
high and variable polarization, and superluminal motion. One of the main defining feature of the blazar emission
is the variability: the flux can increase by orders of magnitude on minute timescales \citep[e.g.,][]{2016ApJ...824L..20A,2018ApJ...854L..26S,2014Sci...346.1080A} or even seconds \citep[e.g.,][]{2007ApJ...664L..71A}.
Even if the emission from blazars is predominantly stochastic, a few sources have shown quasi-periodic oscillations in the \gray\ band
\citep[e.g.,][]{2015ApJ...813L..41A, 2020ApJ...896..134P,2023A&A...672A..86R, 2024ApJ...976..203A}, whose
statistical significant detection remains an ongoing challenge.

The broadband emission of blazars spans from the radio to the high-energy (HE; $>100$ MeV) and very-high-energy
(VHE; $>100$ GeV) $\gamma$-ray band, and exhibits a characteristic two-hump structure in
the $\nu F_{\nu}$ representation \citep[for a review, see e.g.][]{2017A&ARv..25....2P}. The first component peaks between the
IR band and the X-ray band. It is generally interpreted as synchrotron radiation
from relativistic electrons. The second, HE component (peaking above the X-ray band) can be produced from leptonic
or hadronic processes. In leptonic scenarios, it can be produced via inverse Compton scattering—either of synchrotron
photons \citep[SSC;][]{1985A&A...146..204G, 1992ApJ...397L...5M, 1996ApJ...461..657B} or of external photon fields
such as photons from the accretion disk \citep{1992A&A...256L..27D, 1994ApJS...90..945D}, reflected from the
broad-line region \citep{1994ApJ...421..153S}, or from the dusty torus \citep{2000ApJ...545..107B}. Instead, in hadronic
interpretations, the HE component is attributed to proton-synchrotron emission \citep{2001APh....15..121M} and/or
cascades from photo-pion and photo-pair interactions \citep{1993A&A...269...67M, 1989A&A...221..211M, 2001APh....15..121M,
mucke2, 2013ApJ...768...54B, 2015MNRAS.447...36P, 2022MNRAS.509.2102G}. The hadronic (specifically lepto-hadronic) models
have become increasingly attractive following the detection of VHE neutrinos from the direction of blazars, including TXS 0506+056
\citep[e.g.,][]{2018Sci...361..147I, 2018Sci...361.1378I, 2018MNRAS.480..192P} and PKS 0735+178 \citep[e.g.,][]{2023MNRAS.519.1396S} —as well as after statistical studies linking blazars and neutrino events \citep[e.g.,][]{2020ApJ...894..101P,2021MNRAS.504.3338P} . These detection allowed for multimessenger studies of these sources which allowed to better constrain the processes taking place in their jets.

In general, blazars are commonly classified into two groups by the strength of their optical emission lines: BL Lacertae
objects (BL Lacs), which show weak or absent lines, and flat-spectrum radio quasars (FSRQs), which display strong broad
lines \citep{1995PASP..107..803U}. Another classification of blazars is based on the frequency of the synchrotron peak $\nu_p$,
dividing sources into low-synchrotron-peaked (LSP/LBL, $\nu_{p}<10^{14}$ Hz), intermediate-synchrotron-peaked
(ISP/IBL, $10^{14}<\nu_{p}<10^{15}$), and high-synchrotron-peaked (HSP/HBL, $\nu_{p}>10^{15}$ Hz) blazars \citep{Padovani1995, Abdo_2010}. 

The origin of the broadband emission in blazars and the physical changes in jets that drive flaring
activities are active research problems. Although the quantity and quality of the observed data is increasing and the theoretical
models are becoming more sophisticated, time-resolved blazar SED modeling is still limited. To the exception of a few works, see e.g. \citet{2021MNRAS.504.5074S, 2022MNRAS.513.4645S, 2022MNRAS.517.2757S, MohanaA2025, 2025MNRAS.540..582H}, most studies fit an average state, a
single-epoch or at most a few snapshot SEDs.
This is largely due to \textit{(i)} the effort required to assemble strictly contemporaneous broadband data for a given source—often
involving reprocessing and homogenization of heterogeneous archives—and \textit{(ii)} the computational cost of time-dependent
radiative models that track particle injection and cooling while exploring high-dimensional parameter space.
When time-dependent models are coupled to rigorous inference (e.g., full posterior sampling rather than local optimization),
fitting many SEDs becomes numerically intensive. 

To address these issues, we recently developed two complementary tools: \textit{(i)} the
Markarian Multi-wavelength Data Center (\texttt{MMDC}; \citealt{2024AJ....168..289S}), a novel data center that enables access
and retrieval of multi-wavelength and multi-messenger blazar data, including both data from various catalogs and newly analyzed
data in the optical/UV, X-ray, and \gray\ bands; and \textit{(ii)} convolutional neural-network surrogate models trained on synthetic
SEDs that include SSC \citep{2024ApJ...963...71B}, EIC \citep{2024ApJ...971...70S}, and hadronic models \citep{Sahakyan2025}. This promising machine learning approach was also used by \citet{TVP24} in the context of blazars and by \citet{BvL23} for gamma-ray bursts. These surrogates reproduce the radiative signatures of electrons and protons in the jet and can be
coupled with Bayesian inference to fit SEDs efficiently without long computations. Together, these two tools streamline the
construction of time-resolved datasets from blazar observations and enable scalable modeling of large number of SEDs, improving
constraints on the physical processes in blazar jets.

To investigate the origin of blazar-jet emission and its time-dependent evolution, we initiated the Modeling of
time-resolved Spectral Energy Distributions of blazars (MSED) project, which uses the two tools mentioned above.
As a first case study, we analyzed blazar OJ 287, assembling and modeling 739 quasi-simultaneous SEDs within a
one-zone SSC model \citep{2025MNRAS.540..582H}. This enabled to track the temporal evolution of key model parameters
and to identify distinct emission states characterized by specific parameter combinations (magnetic field  $B$, electron index  $p$, and Doppler boost $\delta$).
The second target we studied within the MSED project is the subject of this paper, namely \source, a HSP blazar at redshift $z=0.117$
\citep{1984ApJ...278L.103B}. \source\ is one of the brightest HBL objects and shows pronounced variability across different bands.
For example, hour-scale variability has been observed in the X-ray band \citep[e.g.,][]{2000ApJ...528..243K}, and minute-scale
variability has been observed at VHE \gray\ band \citep{2007ApJ...664L..71A}. This source has been frequently monitored in
different bands, yielding a large, well-sampled dataset, which makes \source\ an ideal target for time-resolved studies of
the broadband emission during different activity states using (quasi-)simultaneous multi-band data.

In this paper, we investigate the physical processes at work in the jet of \source\ by modeling its broadband emission across
different emission states. We use optical/UV, X-ray, and \gray\ observations and construct a large set of SEDs from quasi-simultaneous measurements. Each SED is fitted with a one-zone SSC model, and we analyze the distributions of the physical
parameters to quantify their evolution across states. This allowed us
to quantify the changes in the jet properties responsible for the variation of the emission. The paper is organized as follows.
Section~\ref{analysis} describes the data sets, their reductions, and the light-curve and spectra creation. Section~\ref{sed_evol} presents the
time-resolved SED construction and its temporal evolution. Section~\ref{theory} summarizes the theoretical model and inference
setup. Section~\ref{resdis} reports statistical analysis of the parameters obtained from the modeling in different states. Section~\ref{conc} provides the conclusions.

\section{Data Analysis}
\label{analysis}

In investigating the multiwavelength emission of \source, we used data from \texttt{MMDC} which is an open-access platform that
provides time-resolved SEDs for blazars, combining archival catalog data with newly processed observations in the optical/UV,
X-ray, and \gray\ bands \citep{2024AJ....168..289S}. We note that the data can also be downloaded from the \textit{firmamento} platform \citep{2024AJ....167..116T}. The \texttt{MMDC} platform also enables SED modeling within
leptonic models \citep[SSC and EIC;][]{2024ApJ...963...71B, 2024ApJ...971...70S}—as well as hadronic and lepto-hadronic models \citep{Sahakyan2025}.
From \texttt{MMDC}, both SEDs and multi-band light curves are used to investigate the temporal evolution of the emission from
\source. Below we present the analysis method applied to generate the data and summarize the main results; for more details on
the analysis methods, we refer to \citet{2024AJ....168..289S}.

\begin{figure*}
    \centering
    \includegraphics[width=0.98\linewidth]{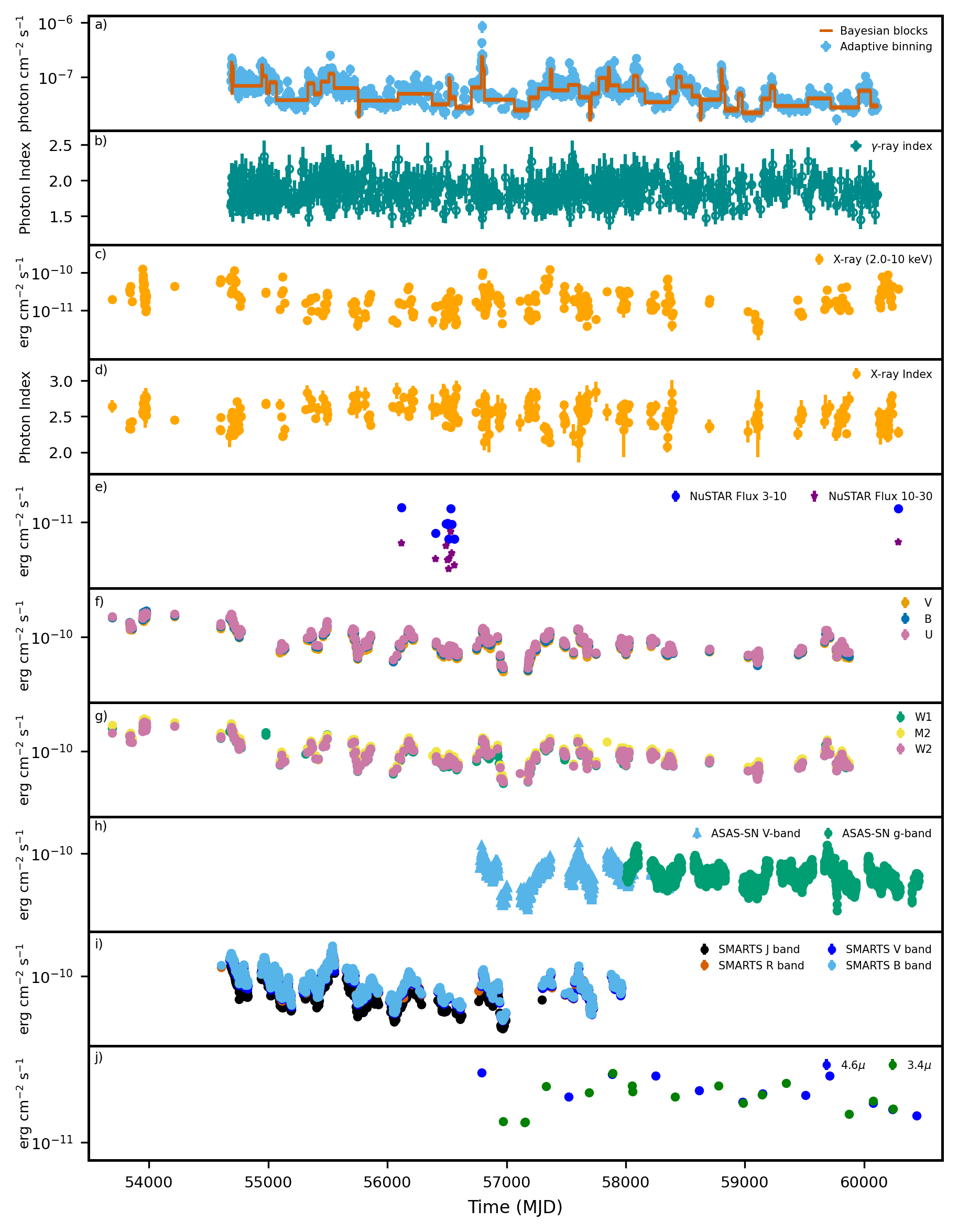}
    \caption{Multiwavelength light curves of PKS 2155-304. \textit{(a)} \gray\ light curve (light-blue adaptive bins) with Bayesian-block segmentation (orange step curve) used to define time-resolved SED intervals. \textit{(b)} \gray\ photon index. \textit{(c)} \textit{Swift}/XRT $2-10$ keV flux. \textit{(d)} Photon index in different XRT observations. \textit{(e)} NuSTAR fluxes in $3-10$ keV (blue diamonds) and $10-30$ keV (magenta pentagons). \textit{(f)} \textit{Swift}/UVOT $U$, $B$, $V$ fluxes. \textit{(g)} \textit{Swift}/UVOT $W1$, $M2$, $W2$ fluxes. \textit{(h)} ASAS-SN $g$- and $V$-band fluxes. \textit{(i)} Fluxes in $J$, $V$, $R$, $B$ bands from SMARTS observations. \textit{(j)} NEOWISE $W1\,(3.4\,\mu\mathrm m)$ and $W2\,(4.6\,\mu\mathrm m)$ fluxes (weighted means over $\leq10$ days windows).
}
    \label{fig:lc}
\end{figure*}

\subsection{$\gamma$-ray data}

Since the launch of the \textit{Fermi} Large Area Telescope (\fermi) in 2008, the \gray\ emission of \source\ has been monitored
continuously. \fermi\ is a pair-conversion \gray\ telescope sensitive to photons from 20~~MeV to 300~~GeV and, in survey mode,
scans the entire sky approximately every three hours. A full description of the instrument is provided by \citet{2009ApJ...697.1071A}.

The analysis follows the standard \fermi\ point-source procedures and is described in detail in \citet{2024AJ....168..289S}; the
key steps are summarized here. We analyzed data from 2008 August 4 to 2023 July 4 (Mission Elapsed Time 239667417–710178221) using
\texttt{Fermitools} v2.0.8 with the \texttt{P8R3\_SOURCE\_V3} instrument response functions. Events in the 100~~MeV–300~~GeV range
were selected from a region of interest (ROI) of radius $12^{\circ}$ centered on \source\ position (RA $=329.71^{\circ}$, Dec $=-30.22^{\circ}$).
We used \texttt{evclass=128} (SOURCE class) and \texttt{evtype=3} (FRONT+BACK) events, and applied a zenith-angle cut of $<90^{\circ}$
to suppress Earth-limb contamination. The source model included all cataloged \gray\ emitters within $17^{\circ}$ of the target selected
from the 4FGL incremental catalog (DR3; \citealt{2022ApJS..260...53A}). Parameters of sources within $12^{\circ}$ were left free during
the fit, while those outside this radius were fixed to their catalog values. The spectra of the sources were assumed to be identical to those used in the 4FGL-DR3 catalog. The Galactic diffuse emission (\texttt{gll\_iem\_v07}) and
the isotropic component (\texttt{iso\_P8R3\_SOURCE\_V3\_v1}) were adopted with the latest recommended templates. We performed a binned
maximum-likelihood analysis to optimize the spectral parameters.

Using the optimized model, we derived the \gray\ light curve of \source\ with an adaptive-binning approach \citep{2012A&A...544A...6L}.
Instead of regular fixed time step, in this method the time bin width is chosen such that the relative flux uncertainty above the optimal
energy ($E_{\rm opt}=276.9$~MeV for \source) reaches a value of 20\%. Consequently, bins are shorter during bright states and longer
during quiescence. For each interval then an unbinned likelihood analysis was applied assuming a simple power-law spectrum
for \source\ (appropriate for short intervals), applying the same event selections and background components as above. Adaptive binning
has been widely used to identify and characterize flaring activity in blazars (e.g., \citealt{2013A&A...557A..71R,2016ApJ...830..162B,2017MNRAS.470.2861S,2017A&A...608A..37Z,2017ApJ...848..111B,2018ApJ...863..114G,2018A&A...614A...6S,2021MNRAS.504.5074S,2022MNRAS.517.2757S,2022MNRAS.513.4645S}).

The \gray\ light curve of \source\ is shown in Figure~~\ref{fig:lc} (a); the corresponding photon index, $\Gamma$, is shown
in Figure~~\ref{fig:lc} (b). The light curve exhibits several episodes in which the \gray\ flux rises to $\sim3\times$ the
long-term mean of $7.32\sim10^{-8}\mathrm{photon\:cm^{-2}\:s^{-1}}$, e.g., near MJD 54693.5, 55517.8, 56789.0–56794.0, and
58089.6. The most prominent \gray\ outburst occurred in May–June 2014: it began around MJD~~56783 (2014-05-05), peaked at
MJD~~56794.7 (2014-05-17), and declined by MJD~~56817 (2014-06-09), for a duration of $\approx34$~~days. At the peak, the
flux reached $8.44\sim10^{-7}\mathrm{photon\:cm^{-2}\:s^{-1}}$ ($>276.9$ MeV) with a very hard photon index of $\Gamma\approx1.47$.
During this flare the spectrum hardened as the flux increased and then softened toward $\Gamma\sim2.0$ during the decay. Similar
episodes—flux enhancements accompanied by spectral hardening— are observed in other periods as well (see Figure \ref{fig:lc} a and b),
indicating that this behavior is common for \source\ in the \gray\ band.

\subsection{X-ray data}

In the X-ray band, \source\ has been regularly monitored by XRT on board
the Neil Gehrels Swift Observatory \citep[][hereafter Swift]{2004ApJ...611.1005G} in the 0.3–10 keV band
and with the Nuclear Spectroscopic Telescope Array \citep[\textit{NuSTAR};][]{2013ApJ...770..103H} in the
3–79 keV band. Their joint coverage enables soft-to-hard X-ray investigation of \source\ emission over more
than two orders of magnitude in energy.

\subsubsection{Swift XRT}
\textit{Swift}/XRT made 343 observations of \source\ until July 4 2023. All blazar observations performed by \textit{Swift} have been analyzed and processed, following the method detailed in \citet{2024AJ....168..289S} and are made publicly available through the \texttt{MMDC} platform \citet{2024AJ....168..289S}.
Here we summarize main steps relevant to reducing the observational data. The data have been processed with the
\texttt{swift\_xrtproc} pipeline \citep{2021MNRAS.507.5690G}, which automatically downloads observation and calibration files, runs
\texttt{xrtpipeline} to generate cleaned event files and exposure maps for each snapshot and the full observation, and performs standard
analysis. Source and background spectra are extracted using a 20-pixel source aperture when the observation is not piled up and a
source-free annulus for the background; pile-up is explicitly checked and, when present, spectra are re-extracted from an annulus
with the inner radius set by the measured count rate. Ancillary response files are built with \texttt{xrtmkarf} and appropriate
response matrices from CALDB. Spectra (0.3–10 keV) are fitted in \texttt{XSPEC} \citep{1996ASPC..101...17A} with a power law model assuming a fixed Galactic absorption $N_{\rm H}$ (using \textit{phabs} model); best-fit models are then converted to $\nu F_{\nu}$ for
SED construction. Further details on \texttt{swift\_xrtproc} are provided in \citet{2021MNRAS.507.5690G}.

The variation of the X-ray flux (2-10 keV) and photon index ($\Gamma_{\rm X}$) in time are presented in Figure~~\ref{fig:lc},
panels c and d respectively. In the X-ray band, several flaring episodes are evident when the flux exceeds the
long-term mean of $\langle F_{2\text{–}10}\rangle \simeq 2.3\times10^{-11}\mathrm{erg\:cm^{-2}\:s^{-1}}$.
The brightest flare occurred at MJD~~53945 (2006-07-29), when the flux reached $(1.25\pm0.02)\times10^{-10} \mathrm{erg\:cm^{-2}\:s^{-1}}$—a
factor of $\sim5.4$ above the mean—with $\Gamma_{\rm X}=2.56\pm0.01$. The second-highest flux was observed at MJD~~57363
(2015-12-07), $(1.21\pm0.15)\times10^{-10}\mathrm{erg\:cm^{-2}\:s^{-1}}$, with a harder spectrum, $\Gamma_{\rm X}=2.13\pm0.06$.
Overall, $\Gamma_{\rm X}$ spans $\simeq2.1$–$2.9$. The hardest spectrum, $\Gamma_{\rm X}=2.07\pm0.07$ was observed at MJD~~58347
(2018-08-17) with a flux of $(6.5\pm0.9)\times10^{-11}\mathrm{erg\:cm^{-2}\:s^{-1}}$, whereas the softest, $\Gamma_{\rm X}=2.90\pm0.10$,
was measured at MJD~~56578 (2013-10-13) at $(4.7\pm0.9)\times10^{-12}\mathrm{erg\:cm^{-2}\:s^{-1}}$. These show strong spectral
and flux variability in the X-ray band, with episodes of both spectral hardening and softening.

\subsubsection{NuSTAR}

\textit{NuSTAR} with its two focal plane modules, FPMA and FPMB \citep{2013ApJ...770..103H} observed \source\ 11 times; nine
of these observations fall within the analysis window considered here. All data of blazars that were observed with \textit{NuSTAR},
including \source{}, are available via \texttt{MMDC}. Here, we summarize the main points of the analysis.
The data were processed with the automated \texttt{NuSTAR\_Spectra} pipeline \citep{2022MNRAS.514.3179M}, built on \texttt{NuSTARDAS}.
The pipeline \textit{i)} downloads the observations and calibration files; \textit{ii)} generates a 3–20 keV image to derive the count
rate to set extraction radii; \textit{iii)} localizes the source with \texttt{XIMAGE}; and \textit{iv)} runs \texttt{nuproducts}
to create high-level science products. Source and background regions are chosen automatically with radii that scale with the measured
count rate (typical the source radius is $\approx30$–$70\arcsec$ while the background is from an annulus separated by $\gtrsim50\arcsec$).
Then, the spectra are grouped to $\ge1$ count per bin and the data are fitted in \texttt{XSPEC} from 3 keV up to the energies when
the background starts to dominate (often 20–79 keV) using Cash statistics \citep{1979ApJ...228..939C} and assuming a power law model and then computing the fluxes in the 3–10 and 10–30 keV bands.
Finally, X-ray SED points are computed from 3 keV up to the energies where the net source signal exceeds the local background.
For more details see \citet{2022MNRAS.514.3179M}.

In Figure~\ref{fig:lc} panel (e), the variation of the fluxes in the 3–10 and 10–30 keV bands is shown displaying
moderate variability, especially in the hard X-ray band. In the 3–10 keV band the flux changes from
$5.6\times10^{-12}$ to $1.7\times10^{-11}\mathrm{erg\:cm^{-2}\:s^{-1}}$ (mean $\approx10^{-11}\mathrm{erg\:cm^{-2}\:s^{-1}}$) while
in the 10–30 keV band, it varies between $2.0\times10^{-12}$ and $7.2\times10^{-12}\mathrm{erg\:cm^{-2}\:s^{-1}}$
(mean $3.7\times10^{-12}\mathrm{erg\:cm^{-2}\:s^{-1}}$). This corresponds to variability amplitudes of
$\sim3.0$ (3–10 keV) and $\sim3.6$ (10–30 keV). The brightest epochs occur at MJD~56116.7 and 56530.8, where the 3–10 keV flux
reached $(1.67\pm0.17)\times10^{-11}$ and $(1.61\pm0.16)\times10^{-11}\mathrm{erg\:cm^{-2}\:s^{-1}}$, respectively. The \textit{NuSTAR}
photon index spans a relatively narrow range, $\Gamma_{\rm X}=2.58$–3.03 (mean $\langle\Gamma_{\rm X}\rangle\simeq2.80$). The
hardest spectrum, $\Gamma=2.58$, was measured at MJD~56489.965 with 3–10 and 10–30 keV fluxes being $9.4\times10^{-12}$ and
$4.4\times10^{-12}\mathrm{erg\:cm^{-2}\:s^{-1}}$, respectively. The softest spectrum, $\Gamma\simeq3.03$, was observed at
MJD~56506.939 with corresponding fluxes of $9.7\times10^{-12}$ and $2.7\times10^{-12}\mathrm{erg\:cm^{-2}\:s^{-1}}$. The soft
indices in this band indicate that the \textit{NuSTAR} band observes the declining, high-energy tail of the synchrotron component
which is characteristic for HSP blazars.

\subsection{Optical/UV data}

\textit{Swift}/UVOT observed \source\ in six filters in the optical (V: 500–600 nm; B: 380–500 nm; U: 300–400 nm)
and ultraviolet (UVW1: 220–400 nm; UVM2: 200–280 nm; UVW2: 180–260 nm) bands providing information on the low energy component of the emission spectrum. UVOT data for all blazars observed by Swift (including \source{}) are accessible via
\texttt{MMDC}. Here, we summarize only the main analysis steps. Data reduction was performed using an automated
pipeline that follows standard UVOT analysis procedures, with photometry extracted from a $5''$ aperture centered on the target and
the background estimated from nearby source-free regions (from a $20''$ region). Each image was inspected for contamination, and
calibrated magnitudes were converted to fluxes using established zero-points and corrected for Galactic extinction based on $E(B-V)$
values from the Infrared Science Archive. For further details, see \citet{2024AJ....168..289S}.

The optical and ultraviolet light curves are shown in Figure~\ref{fig:lc} (panels e and f). Variability is detected in all
\textit{Swift}/UVOT filters, with the source alternating between flaring and quiescent states. The highest flux of
$3.13\times10^{-10}\mathrm{erg\:cm^{-2}\:s^{-1}}$, was observed in \(\mathrm{M2}\) band on MJD $53960.85$, exceeding
the time-averaged flux in the same filter, $1.18\times10^{-10}\mathrm{erg\:cm^{-2}\:s^{-1}}$, by a factor of $2.7$. Among
the remaining filters, the highest flux is $2.78\times10^{-10}\mathrm{erg\:cm^{-2}\:s^{-1}}$ in $\mathrm{W2}$ filter on
the same epoch, which is $2.6$ times higher then the time-averaged flux in this band $1.05\times10^{-10}\mathrm{erg\:cm^{-2}\:s^{-1}}$.
Across the UVOT bands, the ratio of the highest and lowest fluxes ($F_{\max}/F_{\min}$) are $V=8.15$, $B=7.90$, $U=7.04$, $W1=8.29$,
$M2=8.27$, and $W2=7.99$, indicating comparable but band-dependent variability.

In the optical band, we include additional data from the All-Sky Automated Survey for Supernovae
\citep[ASAS-SN;][]{2017PASP..129j4502K} in the $V$ and $g$ filters. The corresponding light curve is shown
in Figure~\ref{fig:lc} panel h. The ASAS-SN light curve exhibits variability consistent with \textit{Swift}/UVOT observations. The mean flux over the monitoring period is $5.8\times10^{-11}\,\mathrm{erg\,cm^{-2}\,s^{-1}}$.
The brightest epoch occurred at MJD 57604.14, when the flux reached $1.31\times10^{-10}\,\mathrm{erg\,cm^{-2}\,s^{-1}}$;
the faintest state was at MJD 59771.34 with the flux of $2.37\times10^{-11}\,\mathrm{erg\,cm^{-2}\,s^{-1}}$, corresponding
to ratio of $\sim5.5$ between the highest and lowest fluxes. These long-term optical measurements show 
pronounced variability of \source\ emission in the optical band.

\subsection{Archival and VHE \gray\ data}

In addition to the data described above, we retrieved multi-band observations of \source\ from \texttt{MMDC}. Some of
these datasets are from catalogs that lack reliable time measurements; therefore these data are used only to construct
the time-averaged SED of \source{}. In contrast, time-resolved datasets—such as data from the Near-Earth Object Wide-field
Infrared Survey Explorer \citep[NEOWISE;][]{2011ApJ...731...53M} and the Small and Moderate Aperture Research Telescope
System \citep[SMARTS;][]{2012ApJ...756...13B} allow to investigate the temporal evolution in the IR and optical bands.
The SMARTS lightcurves in the $J$, $R$, $V$, and $B$ filters are shown in
Figure~\ref{fig:lc}, panel i, demonstrating clear variability in all bands. SMARTS monitoring
shows a major optical flare around MJD 55500: the activity started after MJD 55400 and lasted several weeks. This
flare was detected quasi-simultaneously in all available bands (J, R, V and B). The brightest epoch occurred on
MJD 55538.05, when the $B$-band flux reached $2.34\times10^{-10}\,\mathrm{erg\,cm^{-2}\,s^{-1}}$, a factor of $\sim2.9$
above the long-term average flux from the SMARTS monitoring of \source\ $(\langle F\rangle\simeq8.2\times10^{-11}\,\mathrm{erg\,cm^{-2}\,s^{-1}})$.
Additional flares are present throughout the campaign, but this flare is the most prominent one.
In the infrared band, NEOWISE observed \source\ in different periods at 3.4 and $4.6\,\mu\mathrm{m}$. NEOWISE
observations are organized in short campaigns spanning one to a few days approximately every six months. Therefore, in order
to increase signal-to-noise ration the observations with separation less than 10 days are grouped and combined using a
weighted mean. The resulting light curve with the fluxes grouped around each observations is shown in Figure ~\ref{fig:lc},
panel j revealing moderate variability of the source in the IR band. The maximum weighted-mean flux at
$8.817\times10^{13}\,\mathrm{Hz}$ ($3.4\,\mu\mathrm{m}$) is $(4.57\pm0.02)\times10^{-11}\,\mathrm{erg\,cm^{-2}\,s^{-1}}$
at MJD 56791.00, while at $6.517\times10^{13}\,\mathrm{Hz}$ ($4.6\,\mu\mathrm{m}$) the peak is $(3.81\pm0.01)\times10^{-11}\,\mathrm{erg\,cm^{-2}\,s^{-1}}$
at MJD 56791.03.

\source\ is a well-known emitter in the VHE \gray\ band. To complete the broadband SED, we include published VHE
observations of \source\ by HESS from \citet{2009ApJ...696L.150A} and \citet{2020A&A...639A..42A} and incorporate
them into our dataset. When contemporaneous multi-band coverage exists, these points are included in the time-resolved
SEDs. The VHE data provide crucial information at the highest energies, helping to constrain the second component in
the SED of \source{}.

\begin{figure}
    \centering
    \includegraphics[width=0.48\textwidth]{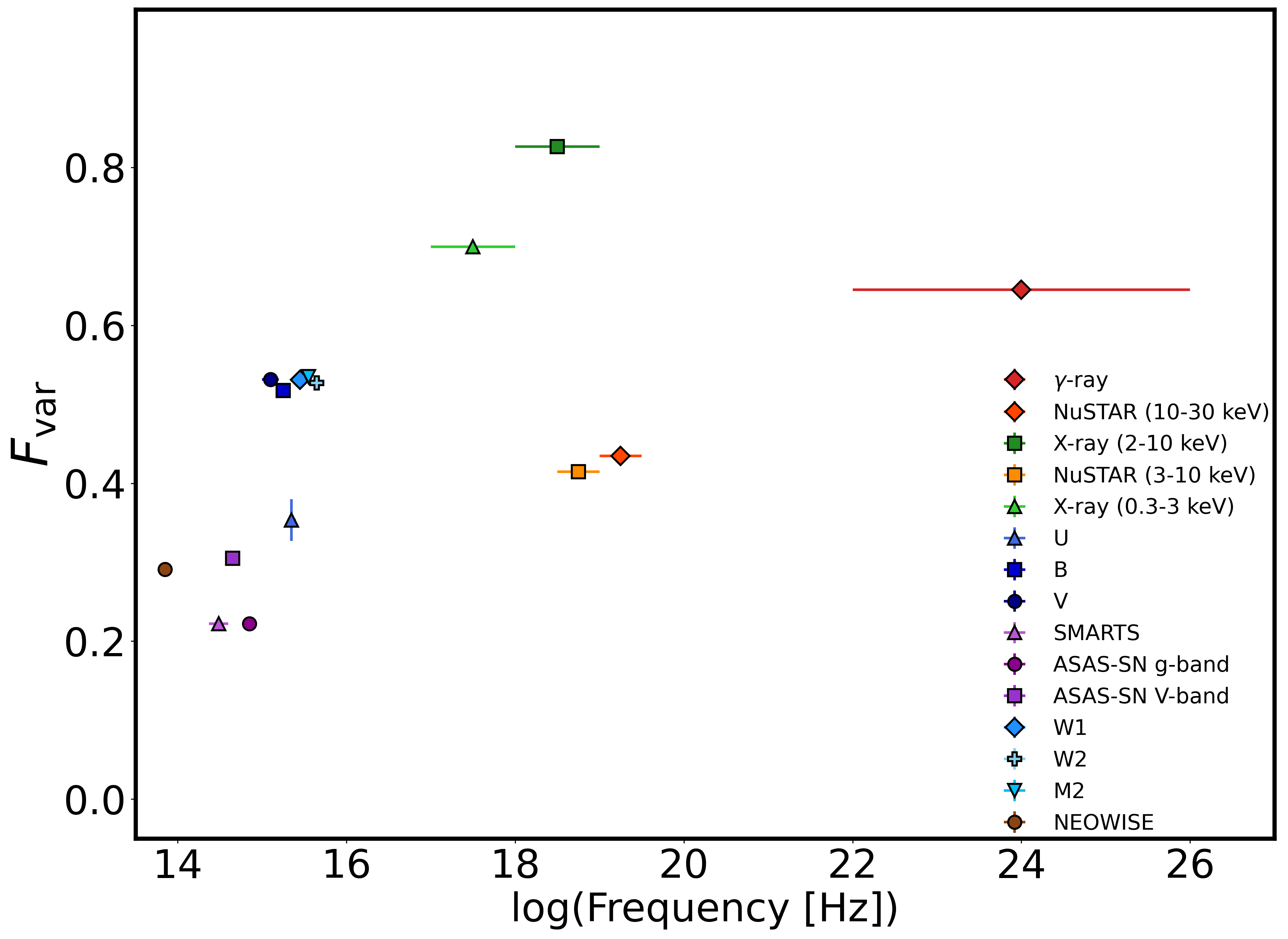}
    \caption{Fractional variability amplitude $F_{\rm var}$ as a function of frequency for \source. Symbols/colours denote bands as in the legend. Vertical error bars are $1\sigma$ uncertainties on $F_{\rm var}$; horizontal bars indicate the width of each band. }
    \label{fig:frac}
\end{figure}

\subsection{Fractional variability}
In order to quantify the variability in different bands, we use the fractional variability $F_{\rm var}$ estimated following \citet{2019Galax...7...62S}. Figure~\ref{fig:frac} shows $F_{\rm var}$ as a function of frequency; horizontal bars indicate the frequency/bandwidth of each instrument/filter, and vertical bars show $F_{\rm var}$ and its uncertainty. The variability amplitude is strongly energy-dependent. In the infrared–optical bands (NEOWISE, SMARTS, ASAS-SN) the variability is low to moderate, $F_{\rm var}\sim0.2\text{--}0.3$. In the optical/UV bands (\textit{Swift}/UVOT) the variability amplitude is higher, $F_{\rm var}\sim0.35\text{--}0.55$, showing more active states in the light curve. The variability is highest in soft/medium X-rays: $F_{\rm var}\sim0.7$ in $0.3\text{--}3$ keV and $\sim0.8$ in $2\text{--}10$ keV bands. At harder X-rays (NuSTAR, $3\text{--}10$ and $10\text{--}30$ keV) the amplitude decreases to $F_{\rm var}\sim0.42\text{--}0.44$. In the HE \gray\ band the variability is moderate with $F_{\rm var}\approx0.65$ (Figure~\ref{fig:frac}).

This trend of fractional variability is overall expected for a blazar of type HSP.
In the spectrum of this type of blazars, the emission in the infrared–optical bands are dominated by low-energy electrons with longer synchrotron cooling times, so the variability amplitude is shorter. Variability grows toward the soft/medium X-rays, which corresponds to the highest-energy tail of the synchrotron component; here acceleration and cooling timescales are shortest, so small changes in the injected electron population, magnetic field $B$, or Doppler factor $\delta$ produce large flux changes. The moderate GeV-band variability is consistent with SSC-dominated inverse-Compton emission in HSPs: the IC component reflects the changes in the same electron population that powers the variability in the X-ray band, but the longer binning partially smooth out the rapid fluctuations.

\section{Time evolution of the multiwavelength SED}
\label{sed_evol}

The datasets assembled above provide a comprehensive, multi-epoch view of the emission from \source\ across the electromagnetic spectrum. This allows to build both a time-averaged SED and a series of time-resolved SEDs that track spectral evolution in different periods. For blazars, this distinction in time-resolved SEDs is essential: variability can shift the normalization and peak frequencies of both the synchrotron and inverse-Compton components. This combination of broad spectral coverage and temporal resolution in different periods is therefore critical to understand the origin of the emission and its changes in time.

In order to build the time resolved SED of \source\, we adopt the SED/LC animation methodology used previously to study the source
BL Lac \citep{2022MNRAS.513.4645S}, 3C 454.3 \citep{2021MNRAS.504.5074S}, CTA 102 \citep{2022MNRAS.517.2757S}, and OJ 287
\citep{2025MNRAS.540..582H}. To build the SED/LC animation, we merge the multi-band light curves in Figure~\ref{fig:lc} with additional
observational data from \texttt{MMDC}, then group the data that are simultaneous or quasi-simultaneous considering the \gray\ intervals
as a base to build the time-resolved SEDs. Namely, the \gray\ light curve is segmented into smaller intervals with the Bayesian Blocks algorithm to
define intervals of approximately constant flux (orange line in Figure~\ref{fig:lc}, panel a). For each block, a dedicated spectral
analysis of \fermi\ data is then performed with the method previously described in Section \ref{analysis} to produce energy-resolved flux points
($5$ or $7$ flux points, depending on the detection significance of the source in the considered interval). Data from all other bands that
are within the Bayesian block window are then binned within the same block applying an adaptive time scan
that subdivides the block into shorter windows to maximize simultaneity of the data while retaining sufficient data points. The resulting
SEDs are plotted and changed sequentially—each aligned with its corresponding light-curve segment—to produce an SED/LC animation that
tracks changes in the emission components. More implementation details on building SED/LC animation are provided in
\citet{2024AJ....168..289S}.

The resulting SED/LC animation for \source\ is available on Youtube at \href{https://youtu.be/nHYfChQSOPU}{\nolinkurl{SED/LC animation}}.
It demonstrates the temporal evolution of the synchrotron and inverse-Compton components and enables to identify spectral component
changes, peak-frequency shifts, flaring periods, etc. The broadband SED of \source\ shows the traditional two-component structure,
with the synchrotron peak at $\nu_{\rm s}\simeq(1\text{–}5)\times10^{16}\,\mathrm{Hz}$ (UV–soft X-ray). Although the X-ray
flux varies substantially, the synchrotron peak frequency remains stable within uncertainties; variability is dominated by
changes in flux with only mild curvature variations and no systematic shift in $\nu_{\rm p}$. The variability is energy-band
dependent and is consistent with the fractional-variability pattern: variability is small to moderate in the optical/UV,
largest in the soft X-ray band, and moderate in the HE \gray\ band.

\section{Modeling of \source\ SEDs}
\label{theory}

\subsection{The SSC model}
\label{sec:modeldescr}

To model the broadband SEDs of \source{}, we adopt a one-zone leptonic synchrotron/SSC model
\citep{1985A&A...146..204G, 1992ApJ...397L...5M, 1996ApJ...461..657B}. In this model, a spherical
emission region of radius $R$ moves relativistically along the jet with bulk Lorentz factor $\Gamma$ and
is viewed at a small angle $\theta \sim 1/\Gamma$ with respect to the jet axis. The region contains a homogeneous
magnetic field of strength $B$ and a non-thermal population of relativistic electrons. The low-energy
component of the SED, extending from radio to X-ray frequencies, is explained by synchrotron radiation
from these electrons as they spiral in the jet magnetic field. The same population upscatters the synchrotron
photons to HEs via the inverse Compton process, producing the HE component. In the current study, the model
is computed using  a novel neural-network–based method presented in \citet{2024ApJ...963...71B}.  In this
approach, the computationally expensive radiative calculations, including electrons cooling, are
replaced by a CNN surrogate model, trained on a large set of synthetic SEDs generated with 
\texttt{SOPRANO} \citep{2022MNRAS.509.2102G} over a wide range of model parameters. \texttt{SOPRANO}
self-consistently solves the coupled kinetic equations for electrons and photons—including particle injection
and radiative cooling—and computes the corresponding SEDs. Once trained, the CNN accurately reproduces the
radiative output of particles inside the jet while reducing the evaluation time by orders
of magnitude, enabling fitting and parameter inference.

In this paper, the injection spectrum of the electron energy distribution is modeled
as a power law with an exponential cut-off,
\begin{align}
\dot Q = \left \{ 
\begin{aligned}
    & Q_{e,0} \gamma^{-p} \exp \left( -\frac{\gamma}{\gamma_{\rm max}} \right) & & \gamma > \gamma_{\rm min} \\
    & 0 && {\rm otherwise}
\end{aligned}\right.
\end{align}
where $Q_{e,0}$ is the normalization such that the electron luminosity is $L_e = \pi R^2 \delta^2 m_e c^3 \int_1^{\infty} \gamma Q_e d\gamma$ , $p$ is the electron power-law index, $\gamma_{\rm min}$ is the minimum electron Lorentz factor at injection (electrons may cool to smaller Lorentz factors), and $\gamma_{\rm max}$ is the maximum electron Lorentz factor. In addition, both electrons and photons are assumed to leave the emitting region in the characteristic dynamical time $t = R/\Gamma$. To conclude, the model has seven free parameters: $p$, $\gamma_{\rm min}$, $\gamma_{\rm max}$, $B$, $R$, $\delta$, and the electron luminosity $L_{\rm e}$, although in practice we do not fit for $\gamma_{\rm min}$, see below.

\subsection{Fit method and classification}

\begin{table*}
\centering
\begin{tabular}{lccccccc}
\hline
State & $B$ [$10^{-2}$ G] & $L_e$ [$10^{44}$ erg s$^{-1}$] & $\gamma_{\rm max}$ [$10^{5}$] & $R$ [$10^{17}$ cm] & $\delta$ & $p$ & $L_B$ [$10^{43}$ erg s$^{-1}$] \\
\hline
QS       & $1.62 \pm 0.02$ & $9.14 \pm 0.84$ & $2.09 \pm 0.01$ & $3.31 \pm 0.37$ & $24.6 \pm 6.1$ & $2.30 \pm 0.21$ & $6.55 \pm 3.19$ \\
MWF & $4.78 \pm 0.19$ & $5.36 \pm 1.69$ & $4.87 \pm 0.30$ & $1.88 \pm 0.11$ & $24.8 \pm 5.2$ & $2.20 \pm 0.13$ & $18.5 \pm 6.6$ \\
$\gamma$F      & $1.71 \pm 0.17$ & $13.7 \pm 8.5$  & $1.83 \pm 0.00$ & $4.64 \pm 0.47$ & $18.2 \pm 4.9$ & $2.06 \pm 0.09$ & $7.78 \pm 1.91$ \\
XF          & $1.64 \pm 0.21$ & $6.48 \pm 6.14$ & $9.81 \pm 0.12$ & $5.34 \pm 0.72$ & $18.8 \pm 5.4$ & $2.21 \pm 0.14$ & $10.1 \pm 2.4$ \\
OUF           & $1.28 \pm 0.14$ & $11.1 \pm 8.07$ & $2.33 \pm 0.01$ & $5.63 \pm 0.63$ & $20.8 \pm 5.4$ & $2.17 \pm 0.14$ & $8.42 \pm 2.43$ \\

Unclassified  & $1.56 \pm 1.35$ & $9.42 \pm 6.56$ & $1.51 \pm 0.62$ & $4.53 \pm 3.91$ & $23.0 \pm 5.4$ & $2.25 \pm 0.14$ & $9.98 \pm 2.70$ \\

\hline
\end{tabular}
\caption{Model parameters from one-zone SSC fits to the SEDs of \source\ in the different states shown in Figure~\ref{fig:seds}. These are provided as examples, but each of the 253 SEDs have been fitted and their parameters recovered. }
\label{table:param}
\end{table*}

In the SED/light-curve animation discussed in Section \ref{sed_evol}, a total of 327 SEDs are available. However, not all of
these can be reliably modeled due to insufficient observational coverage. Among them, only 253 SEDs have adequate multiwavelength
coverage enough for the modeling: specifically, we require the availability of X-ray data together with
measurements in the low-energy band, as the \gray\ data is available by default. These SEDs were therefore selected for 
modeling. The fits were performed using MultiNest, a nested sampling algorithm \citep{FHB09}. We
adopted 1500 active points and a tolerance of 0.4, which ensures both efficient exploration of the parameter space and robust
convergence of the posterior distributions. For the modeling, we fixed the minimum Lorentz factor of the electron distribution to
$\gamma_{\rm min} = 100$. This choice is motivated by two considerations. First, the low-energy emission (below $\sim 10^{10}$ Hz)
is often produced in more extended regions of the jet, which are not well constrained by the compact one-zone model applied here.
Second, observational data at low frequencies are not available in most of the SEDs selected here. Setting $\gamma_{\rm min}=100$
avoids introducing unconstrained degrees of freedom while remaining consistent with typical values adopted in blazar modeling.
Finally, we note that extragalactic background light absorption at high energy is included by multiplying the emission model 
described in Section \ref{sec:modeldescr} by the model developed in \citet{2011MNRAS.410.2556D}.

In order to have a quantitative discussion on the emission states of \source{}, the selected SEDs were grouped in different
categories based on the activity states. Namely, the dataset was divided into three frequency ranges: optical/UV
($3 \times 10^{14} - 10^{16}\,\mathrm{Hz}$), X-rays ($10^{16} - 10^{21}\,\mathrm{Hz}$), and \grays\ ($10^{21} - 10^{28}\,\mathrm{Hz}$).
For each range, we computed the average flux and its standard deviation which serve as a baseline level. Then, the state of
the source was classified in the following way. A multiwavelength flare (MWF) was identified when the flux in all three bands exceeded the archival average, whereas a quiescent state (QS) was defined when the flux in all bands is on average or remained below the average. If only the \gray\ flux was above the average while the UV/optical and X-ray bands were below, the episode was classified as a \gray\ flare ($\gamma$F). Similarly, an X-ray flare (XF) corresponded to a case where the X-ray flux was above the average while the other bands remained lower, and a UV/optical flare (OUF) was defined when only the optical/UV flux exceeded the average. A total of 117 SEDs were classified as QS, 18 as MWF, 7 as $\gamma$F, 20 as XF, and 31 as OUF. The remaining 61 SEDs display mixed characteristics in the mentioned bands and are left unclassified. This separation makes it possible to separate true multiwavelength flares from band-limited flares and quiescent states, thereby providing a clearer characterization of the variability pattern of \source\ across the electromagnetic spectrum.

\begin{figure*}
    \centering
    \includegraphics[width=0.98\linewidth]{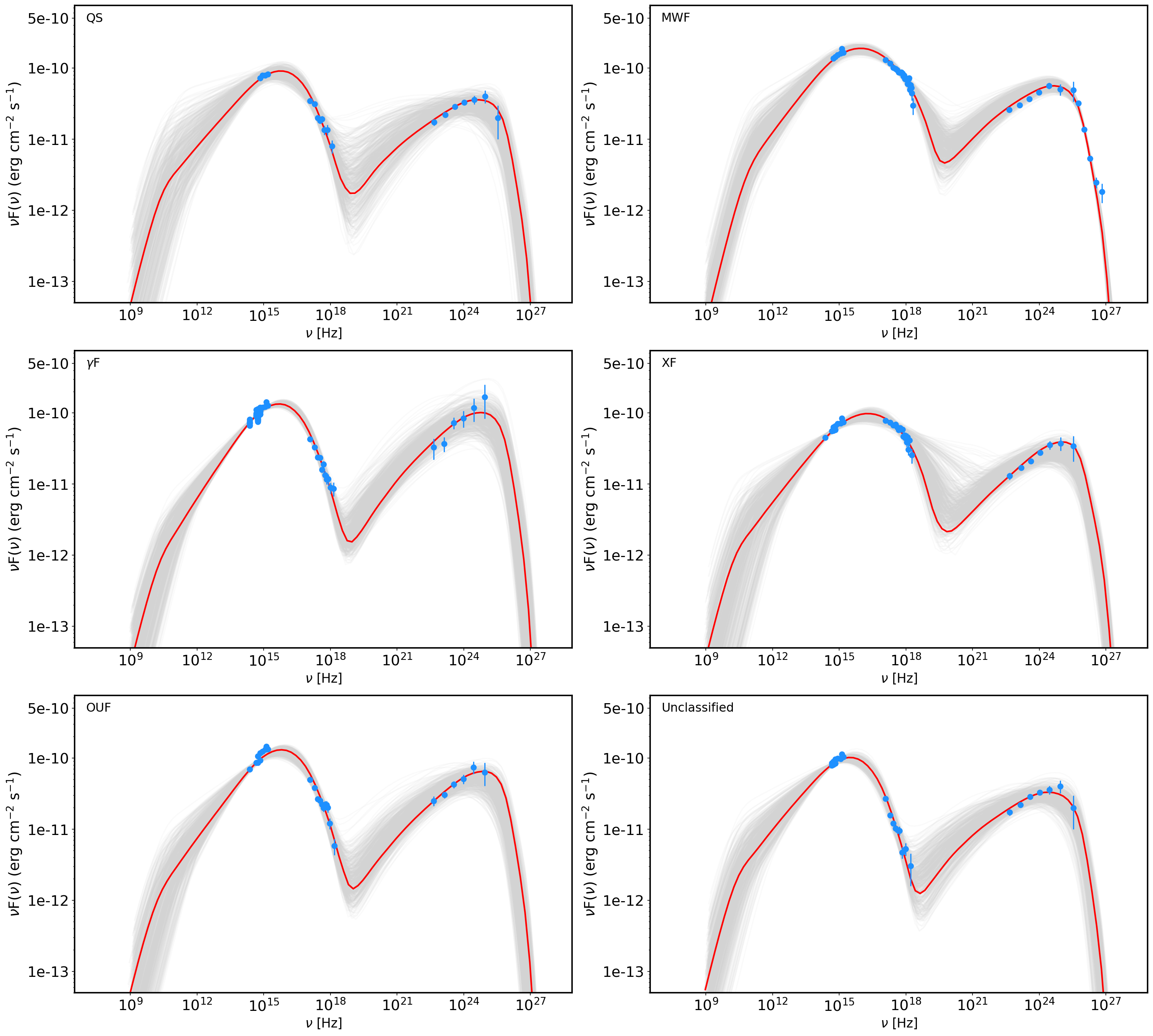}
    \caption{ For each SED class, example of a multiwavelength SEDs of \source\ modeled within the one-zone SSC framework. The blue points are the data, the red curve corresponds to the best-fit model (e.g., when the likelihood is maximum) and the gray spectra corresponds to the model uncertainty. Extragalactic background light absorption is included in the modeling using the model of \citet{2011MNRAS.410.2556D}.}
    \label{fig:seds}
\end{figure*}

\subsection{Example of fit results for each category}
Before studying in detail the statistical properties of each SED category, we provide for each one independently an
example demonstrating the fitting performances and results. 

An example of the SED corresponding to the QS is shown in the upper-left panel of Figure \ref{fig:seds} and the
corresponding parameters are given in Table \ref{table:param}. The red solid line represents the model obtained
with the best-fit parameters, while the gray lines indicate the associated uncertainties. In this state, the
emission can be explained with (i) an electron power-law index $p = 2.3$, (2) a maximum Lorentz factor of the electron distribution $\gamma_{\rm max} = 2.1 \times 10^{5}$, (3) a magnetic field strength $B = 1.6 \times 10^{-2}$ G, (4) an emitting region radius $R = 3.3 \times 10^{17}$ cm, and (5) a Doppler factor $\delta = 24.6$. The total kinetic power carried by electrons is $L_{\rm e} = 9.1 \times 10^{44}$ erg s$^{-1}$, while the power in the magnetic field, computed as $L_{B} = \pi c R^{2} \Gamma^{2} U_{B}$, is $6.5 \times 10^{43}$ erg s$^{-1}$.

The SED corresponding to the MWF is presented in the upper-right panel of Figure \ref{fig:seds} with corresponding
parameters in Table \ref{table:param}. In this case, the emission is produced from a region with a magnetic field
of $B = 4.8 \times 10^{-2}$ G and an electron kinetic power of $L_{\rm e} = 5.4 \times 10^{44}$ erg s$^{-1}$. The
maximum Lorentz factor of the electron distribution is $\gamma_{\rm max} = 4.9 \times 10^{5}$, indicating the presence
of highly energetic particles in the emitting region. As we demonstrate below, these high energy electrons are characteristics of the MWF activity.
The size of the emitting region is $R = 1.9 \times 10^{17}$ cm, and the Doppler factor is $\delta = 24.8$ demonstrating a
significant relativistic boosting during this MWF. The electron energy distribution
is characterized by a power-law index $p = 2.20$. 

The SEDs corresponding to $\gamma$F, XF, and OUF states are shown in the middle and lower-left panels of Figure \ref{fig:seds}.
In all cases, the broadband emission is well modeled with the SSC model used in this study, with the best-fit parameters
for these examples summarized in Table \ref{table:param}. In the $\gamma$F state, the fit yields a magnetic field of
$B = 1.71 \times 10^{-2}$ G, an electron luminosity of $L_{\rm e} = 1.37 \times 10^{45}$ erg s$^{-1}$, and a maximum Lorentz
factor of $\gamma_{\rm max} = 1.8 \times 10^{5}$. The emitting region size is $R \simeq 4.6 \times 10^{17}$ cm, and the Doppler
factor is moderately low compared to other states ($\delta \simeq 18.2$). The electron slope is harder than in the QS ($p = 2.06$),
consistent with harder particle injection during the \gray\ activity. In the XF state, the electron energy distribution reaches
significantly higher cutoff energies, with $\gamma_{\rm max} \simeq 9.8 \times 10^{5}$, almost an order of magnitude larger than
in the  $\gamma$F and OUF states. This shift is accompanied by a relatively weak magnetic field ($B = 1.64 \times 10^{-2}$ G)
and large emission region ($R \simeq 5.3 \times 10^{17}$ cm). The Doppler factor remains moderate ($\delta \simeq 18.8$),
while the electron luminosity is comparatively lower ($L_{\rm e} \simeq 6.5 \times 10^{44}$ erg s$^{-1}$). These values
indicate that the electrons are effectively accelerated to higher energies which drives the strong changes in the X-ray
component. On the contrary, the OUF state is characterized by a higher magnetic field ($B = 2.04 \times 10^{-2}$ G) and a
more compact emission region ($R \simeq 3.5 \times 10^{17}$ cm). The maximum Lorentz factor is the lowest among the flaring
states ($\gamma_{\rm max} \simeq 1.4 \times 10^{5}$), suggesting that the flare is dominated by the emission from low energy
electrons rather than re-acceleration or injection of fresh energetic electrons. The Doppler factor
remains high ($\delta \simeq 23.9$), also in this case.

An example of modeling of the SED which was unclassified is shown in the lower-right panel of Figure~\ref{fig:seds}, and
the corresponding parameters are listed in Table~\ref{table:param}. In this epoch, the optical/UV and \gray\ fluxes are
consistent with their long-term averages, whereas the X-ray flux is comparatively low. In this case a one-zone SSC model reproduces the broadband SED well with parameters that are not substantially
different than those obtained in the other cases.

As seen from these examples, while the SSC model provides a satisfactory description of the
data in all cases, the key differences in the parameters during different states — namely, the high $\gamma_{\rm max}$ in
the XF state, the flatter electron spectrum in the $\gamma$F state, and the stronger magnetic field in the OUF state -
show that distinct physical processes dominate in each activity state. These examples suggest that the spectral variability in \source{} 
is not driven by a single mechanism, but rather by changes in the emitting particle or in the emission region. It is the
purpose of Section \ref{resdis} to demonstrate the statistical differences across the different emission episodes and isolate parameter changes
responsible for changes in the spectrum of \source{}, to enable a physical description of the different emission periods of \source{}. 

\subsection{The previous modelings of \source{}}
In contrast to earlier studies, where the modeling of \source\ was performed for only a limited number of broadband SEDs, the
analysis presented here considers modeling of a substantially larger number of SEDs. The results obtained in this study across
different states are typical of those usually found for blazars and consistent with previous studies of \source. For example,
\citet{2016ApJ...831..142M} modeled the source during a low-flux state using contemporaneous multiwavelength data. Their results
indicated that the emission can be explained with an electron energy distribution with an index $p = 2.2$, a break Lorentz factor
$\gamma_{\rm br} \sim 2.6 \times 10^{4}$, and a magnetic field strength of $B \sim 0.5$ G, with a characteristic emitting region
radius of $R \sim 1.3 \times 10^{16}$ cm. Most of these parameters (except $B$) are within the range that we estimated for the QS,
see Section \ref{resdis}. Another modeling of \source{} was presented in \citet{2012A&A...544A..75A} based on the 2006 MAGIC
campaign; the SED was modeled by adopting (and not fitting) the parameters $\delta = 50$, $B \simeq 0.085,\mathrm{G}$,
and $R \simeq 9\times10^{15},\mathrm{cm}$. However, in their SED the contemporaneous HE $\gamma$-ray data were absent limiting
the constraints provided by the inverse Compton component. The main difference between the results
obtained here and these previous studies of \source\ is that we performed modeling without assuming any specific initial spectral
shape, instead exploring the global minima of the posterior distributions for each SED across many time periods. %

\begin{figure*}
    \centering
    \includegraphics[width=1\linewidth]{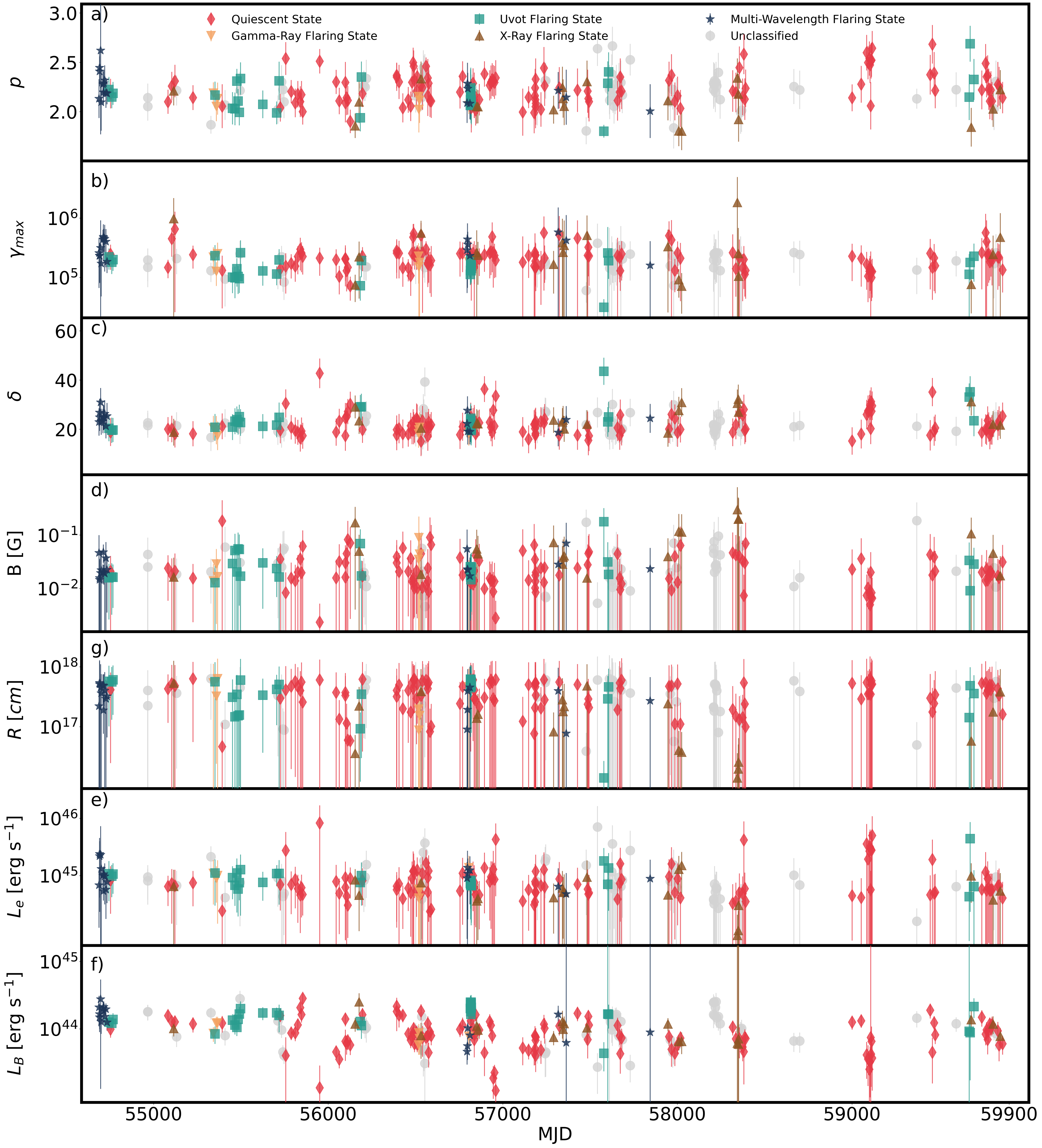}
    \caption{Time evolution of the one-zone SSC parameters for \source{}, derived from fitting the SEDs
    in different states. Panels show (a) the electron power-law index $p$, (b) the maximum electron Lorentz factor
    $\gamma_{\rm max}$, (c) the Doppler factor $\delta$, (d) the magnetic field $B$ [G], (e) the emitting-region
    radius $R$ [cm], (f) the electron kinetic power $L_{e}$ [erg s$^{-1}$] , and (g) the (derived) magnetic power
    $L_{B}$ [erg s$^{-1}$] versus MJD. Colored symbols denote activity states as indicated in the legend.}
    \label{fig:param_lc}
\end{figure*}

\section{Statistical analysis of the different states}
\label{resdis}

Multiwavelength modeling plays a central role in advancing our understanding of blazar emission, especially when different states can be compared. This approach allows to compare and contrast key model parameters,
which in turn provides a direct link between the observed variability patterns and the underlying physical conditions in the jet.
In this way, observational properties, e.g., the flux changes or spectral variability in different bands, can be connected with physical parameters characteristics of the emission region, such as the spectral slopes of the emitting particles, the
magnetic field, the maximum energy of the electrons, etc. These can then be transformed into constraints
on particle acceleration, energy dissipation, and radiation processes.

The modeling of the selected 253 time-resolved SEDs together with the
time evolution of the posterior distributions of the model parameters is presented as an animation at
\href{https://youtu.be/jqssDn5Gjs8}{\nolinkurl{YouTube}} synchronized with the corresponding light-curve intervals, connecting
the evolution of both the multi-wavelength emission state and the SED modeling.
The images from the animations are available on \href{https://github.com/gevorgharutyunyan/PKS-2155-304/tree/main/time_resolved_sed_modeling}{GitHub}. The temporal evolution of the parameters is shown in Figure \ref{fig:param_lc}, where different emission states are represented by distinct colors and symbols. Such a representation allows to evaluate changes in parameter space that are responsible for flaring activities when comparing with the light curve from Figure \ref{fig:lc}, either across the full multiwavelength band or within specific energy ranges, thereby offering clues about the physical origin of these events.

\begin{table*}
\centering
\caption{ Results of the KS test, indicating if the parameter variation between any state and the QS are significant. The parameters are in this order:} magnetic field ($B$), electron luminosity ($L_e$), maximum energy ($\gamma_{\rm max}$), emission region size ($R$), Doppler factor ($\delta$), spectral index ($p$).
\begin{tabular}{|l|cc|cc|cc|cc|cc|cc|}
\hline
State & \multicolumn{2}{c|}{$B$} & \multicolumn{2}{c|}{$L_{\rm e}$} & \multicolumn{2}{c|}{$\gamma_{\rm max}$} & \multicolumn{2}{c|}{$R$} & \multicolumn{2}{c|}{$\delta$} & \multicolumn{2}{c|}{$p$} \\
\cline{2-13}
      & KS & $\mathfrak{p}$ & KS & $\mathfrak{p}$ & KS & $\mathfrak{p}$ & KS & $\mathfrak{p}$ & KS & $\mathfrak{p}$ & KS & $\mathfrak{p}$ \\
\hline
MWF        & 0.338 & 0.045 & 0.359 & 0.027 & 0.474 & 1.0$\times10^{-3}$ & 0.252 & 0.236 & 0.410 & 0.007 & 0.137 & 0.899 \\
$\gamma$F  & 0.267 & 0.354 & 0.293 & 0.254 & 0.327 & 0.155 & 0.265 & 0.364 & 0.239 & 0.489 & 0.506 & 0.004 \\
XF         & 0.545 & 3.0$\times10^{-5}$ & 0.301 & 0.072 & 0.338 & 0.030 & 0.519 & 8.8$\times10^{-5}$ & 0.390 & 0.008 & 0.405 & 0.005 \\
OUF        & 0.316 & 0.007 & 0.381 & 4.8$\times10^{-4}$ & 0.418 & 8.9$\times10^{-5}$ & 0.143 & 0.581 & 0.290 & 0.016 & 0.283 & 0.021 \\
\hline
\end{tabular}
\label{tab:ks_comparison}
\end{table*}

\subsection{Statistical properties of the inferred model parameters}

From modeling the 117 SEDs in the QS state we found that, the best-fit model parameters exhibit relatively narrow distributions. The electron
power-law index is centered around a mean of $p \simeq 2.24$ (median 2.22), varying between $1.90$ and $2.68$. The electrons maximum Lorentz
factor spans nearly an order of magnitude, from $\gamma_{\rm max} \simeq 7.2 \times 10^{4}$ to
$6.5 \times 10^{5}$, with an average value of $2.3 \times 10^{5}$. The magnetic field strength is typically weak, ranging from
$B \simeq 2.3 \times 10^{-3}$ to $1.8 \times 10^{-1}$ G, with a mean of $2.6 \times 10^{-2}$ G. The size of the emitting region
is large, with radii between $R \simeq 4.7 \times 10^{16}$ and $6.3 \times 10^{17}$ cm (mean $3.8 \times 10^{17}$ cm). The Doppler
factor varies moderately, with values from $\delta \simeq 15.1$ to 42.7 (mean $\sim 21.9$). The electron kinetic power is of the order
of $L_{\rm e} \sim 10^{45}$ erg s$^{-1}$, ranging from $2.4 \times 10^{44}$ to $8.3 \times 10^{45}$ erg s$^{-1}$, with a mean value
of $1.0 \times 10^{45}$ erg s$^{-1}$.

The modeling of the 18 SEDs identified as MWF shows  that the magnetic field strength
varies within $B \simeq 1.5 \times 10^{-2}$ - $7.0 \times 10^{-2}$ G, with an average value of $2.9 \times 10^{-2}$ G. The electron luminosity varies in the range between $4.8 \times 10^{44}$ erg s$^{-1}$ and $2.4 \times 10^{45}$ erg s$^{-1}$ with a mean of $1.1 \times 10^{45}$ erg s$^{-1}$. The maximum Lorentz factor of the electron distribution is between $1.6 \times 10^{5}$ and $5.8 \times 10^{5}$, with a mean value of $3.4 \times 10^{5}$, which shows the presence of electrons accelerated to GeV energies  during flares. The emission region is found to be relatively extended, $R = 0.8 \times 10^{17}$–$5.4 \times 10^{17}$cm, with a mean radius of $3.4 \times 10^{17}$ cm. The Doppler factors cluster around $\delta \sim 24$, ranging from 18.7 to 30.9, while the power-law index of electron energy distribution changes between $p = 2.0$ and $2.6$ with a mean of 2.23. 

In the $\gamma$F state, the modeling shows that the emission regions are moderately extended, $R \sim 0.9$–$6.3 \times 10^{17}$cm,
with relatively weak magnetic fields, $B \sim 1.3 \times 10^{-2}$–$8.9 \times 10^{-2}$G. The maximum energy of the electron
distribution $\gamma_{\rm max}$ is in the range between $1.0 \times 10^{5}$ and $2.5 \times 10^{5}$, while
the electron luminosity are in the range of $L_{\rm e} \sim (0.4-1.4) \times 10^{45}$ erg.s$^{-1}$ with a mean of
$L_{\rm e} = 8.4 \times 10^{44}$ erg s$^{-1}$. In these periods, the power-law index of the electrons is stepper than in other ones, with
$p=2.02-2.18$, indicating that more energy is present in the highest energy electrons, while the Doppler boost is
in the range $\delta \sim 17-21$.

In contrast, the modeling shows that XF have systematically higher electron maximum energies and magnetic
field: $\gamma_{\rm max}$ varies in the range of $7.1 \times 10^{4}$ and $1.8 \times 10^{6}$ with 70\% of the estimated $\gamma_{\rm max}$
being above  $2 \times 10^{5}$. The magnetic field is stronger on average, varying between $B \sim 1.6 \times 10^{-2}-0.3$ G, while
the emission regions are somewhat more compact, $R \sim 1.4 \times 10^{16}$–$5.3 \times 10^{17}$cm. The modeling of these flares
also results in a high Doppler factor, $\delta \sim 18.5-32.2$, consistent with enhanced beaming
during strong X-ray activity, and the electron luminosity is in the range between $9.0 \times 10^{43}$ erg s$^{-1}$ to $L_{\rm e}=1.5 \times 10^{45}$ erg s$^{-1}$ with a mean of $L_{\rm e} = 6.06 \times 10^{44}$ erg s$^{-1}$. 

The modeling of SEDs in the OUF states shows yet another behavior: the magnetic field is moderate, $B=9.0 \times 10^{-3}-0.18$ G, with a mean of $3.7\times10^{-2}$ G, but
the emission regions are systematically larger, $ R= 1.4\times10^{16}-6.2 \times 10^{17}$cm, with 89\% of the cases having
an emission region size exceeding $10^{17}$cm. The maximum Lorentz factor of the electrons is comparatively low, $\gamma_{\rm max}$
is in the range between $3.1 \times 10^{4}$ and $2.6 \times 10^{5}$ which is similar to the values estimated in the $\gamma$F state
but well below those estimated for XF. The Doppler boost is relatively strong varying in the
interval $\delta = 19.4-43.5$, and the electron luminosity is comparable to the values estimated in $\gamma$F, with
$L_{\rm e}= 0.4-4.4  \times 10^{45}$ erg s$^{-1}$.

\begin{figure*}
    \centering
        \includegraphics[width=0.98\linewidth]{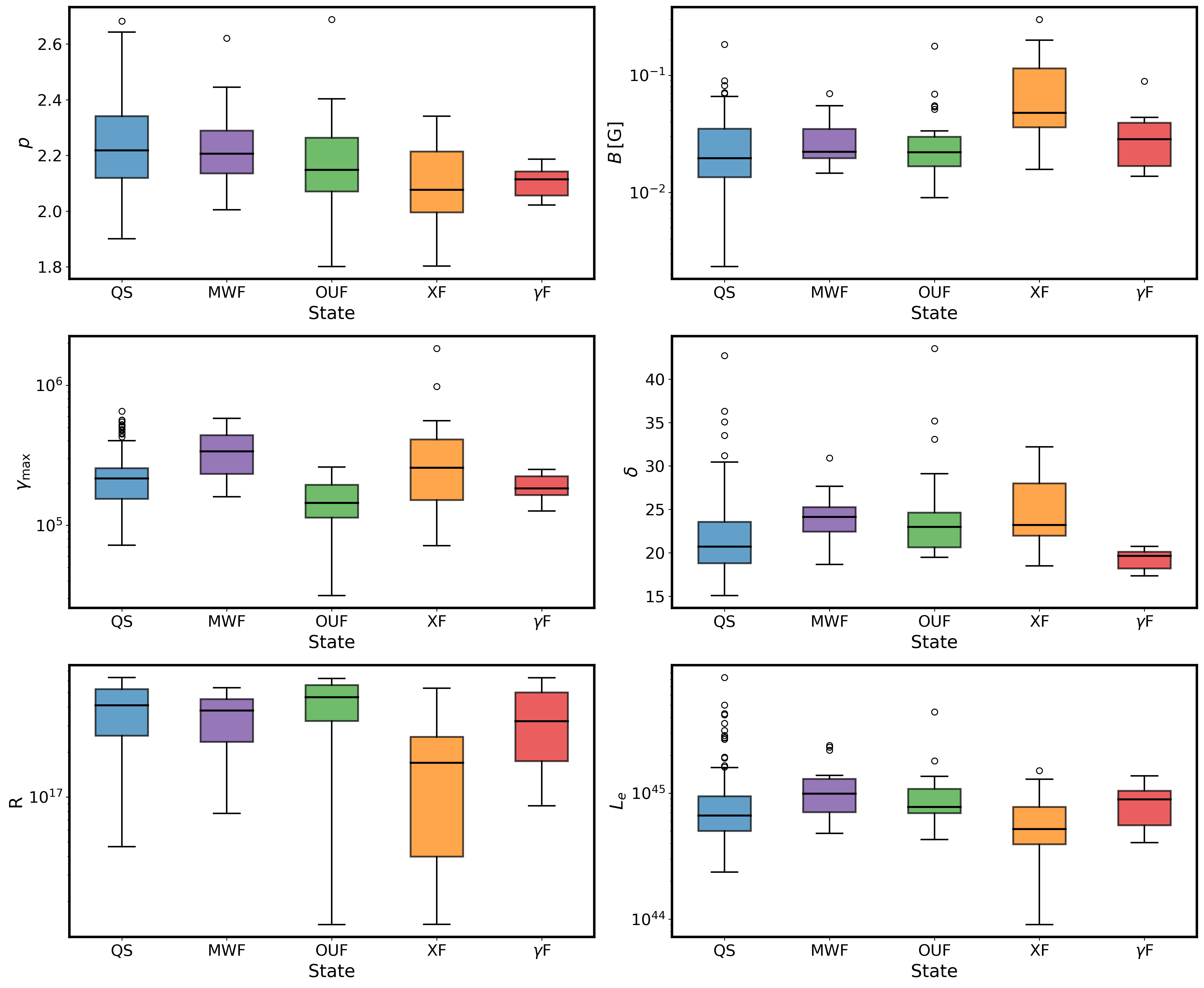}
        \caption{The distributions of all model parameters $p$, $B$ [G], $\gamma_{\rm max}$, $\delta$, $R$ [cm] and $L_e$ [erg.s$^{-1}$]  for different states represented as box plots. The central line marks the median, box edges correspond to the 25th and 75th percentiles, whiskers extend to $1.5\times$ the interquartile range, and outliers are plotted as circles. Qualitatively, the X-ray–flaring state tends to show larger $B$, while $\gamma$-ray–flaring states favor slightly harder $p$ and lower $\delta$ and $\gamma_{\rm max}$ relative to other states.}
        \label{fig:box}
\end{figure*}

To clearly display the differences between the different emission states, we show on Figure \ref{fig:box}, the distributions of the parameters for different states separately. This comparison shows several
general trends. (i) In most cases, the mean of the electron power-law index remains relatively stable around $p \sim 2.1 - 2.2$, with only a
modest hardening during XF and $\gamma$F. In contrast, (ii) the mean of the magnetic field shows a strong dependence on the spectral state: it is weakest in the QS and OUF ($B \lesssim 0.03$ G), moderate during MWF and $\gamma$F, and highest
in XF ($B \gtrsim 0.08$ G), representing a variation of a factor of nearly 3 on average between the different states. The mean of the
Doppler factor clusters around $\delta \sim 20 - 25$ in most cases.
(iii) The mean of the electron maximum Lorentz factor allows to distinguish the states very
clearly: it is the highest in the MWF and XF ($\gamma_{\rm max} \gtrsim 3\times10^{5}$), comparatively
lower in OUF state ($\gamma_{\rm max} = 1.5\times10^{5}$), and moderate in $\gamma$F ($\gamma_{\rm max} = 1.7\times10^{5}$).
(iv) The distribution of $R$ shows relatively extended emission regions across all states, clustering around
a few $\times 10^{17}$ cm. While the median values do not differ drastically, XF exhibits the widest spread, with $R$
reaching both the lowest and highest extremes among the states. In contrast, QS, MWF, OUF, and $\gamma$F tend to show
more confined distributions with median values around $3-4\times10^{17}$ cm. (V) The distribution of $L_{\rm e}$ is
comparatively stable across all states, with mean values around $10^{44.5}-10^{45}\,\mathrm{erg\,s^{-1}}$.
The QS exhibits a slightly broader spread, extending toward higher luminosities, while the flaring states (MWF, XF, and $\gamma$F) appear more clustered and show no systematic enhancement in $L_{\rm e}$. 

In order to investigate differences in the parameter distributions between the QS and other activity states, we performed
Kolmogorov-Smirnov (KS) tests on the parameters. The results are summarized in Table \ref{tab:ks_comparison}, for each
parameter providing the KS statistics and corresponding $\mathfrak{p}$-values (probability). We consider $\mathfrak{p}$-values
below 0.05 (roughly corresponding to 2$\sigma$) as indicating
statistically significant differences from the Quiet State distribution. For the MFW state, significant differences
appear in $B$ ($\mathfrak{p} = 0.045$), $L_{\rm e}$ ($\mathfrak{p} = 0.027$), $\gamma_{\rm max}$ ($\mathfrak{p} = 0.001$),
and $\delta$ ($\mathfrak{p} = 0.007$), suggesting that these parameters change during flares compared to QS. In contrast,
$R$ and $p$ show no notable differences ($\mathfrak{p} > 0.2$), meaning that the emission region size and electron spectrum
remain similar to the QS. In the $\gamma$F state, most parameters do not differ significantly from the QS, with $\mathfrak{p}$-values
above 0.15 for $B$, $L_{\rm e}$, $\gamma_{\rm max}$, $R$, and $\delta$. However, the spectral index $p$ shows a clear
hardening ($\mathfrak{p} = 0.004$), pointing to changes in the electron energy distribution during \gray{} flares. The largest number of parameter changes are observed during the XF state with significant $p$-values
for $B$ ($\mathfrak{p} < 0.001$), $\gamma_{\rm max}$ ($\mathfrak{p} = 0.030$), $R$ ($\mathfrak{p} < 0.001$), $\delta$
($\mathfrak{p} = 0.008$), and $p$ ($\mathfrak{p} = 0.005$). Only $L_e$ does not show significant changes ($\mathfrak{p} = 0.072$),
indicating that X-ray flares involve broad changes in the magnetic fields, emission geometry, beaming,
as well as in the particle distribution, but electron luminosity is comparable to quiescence. For the OUF state, we find
significant differences in all parameters except $R$ ($\mathfrak{p} = 0.581$). For the other parameters, the significance
is given by $B$ ($\mathfrak{p} = 0.007$), $L_{\rm e}$ ($\mathfrak{p} < 0.001$), $\gamma_{\rm max}$ ($\mathfrak{p} < 0.001$), $\delta$
($\mathfrak{p} = 0.016$), and $p$ ($\mathfrak{p} = 0.021$). This suggests that flares in the optical/UV bands are the results
of changes in magnetic fields, electron properties, and beaming, while the emission region size does not deviate much from
the QS.

Overall, these KS test results show state-dependent variations in the physical parameters. The XF and OUF states show
the strongest departures from QS, particularly for $\gamma_{\rm max}$ and $\delta$. This could be due
to enhanced particle acceleration and relativistic effects during flaring activities. The limited changes
in the $\gamma$F state, except $p$, suggest that \gray{} flares might be due to different mechanisms, which affects the
electron spectrum but not the other parameters.

\subsection{Physical interpretation }

The modeling across different periods reveals state-dependent changes in several parameters that inform the emission scenario.
The electron power-law index varies during the $\gamma$F, XF, and OUF states but remains within $p\simeq 1.8$–$2.7$, i.e.,
within the range expected from shock-acceleration theory. Indices near $\simeq 2.0$ are characteristic of diffusive shock acceleration
\citep{1978MNRAS.182..147B,1978ApJ...221L..29B,1987PhR...154....1B}, while softer or harder spectra can arise in relativistic shocks
\citep{1987ApJ...315..425K,1990ApJ...360..702E,1998PhRvL..80.3911B,2004APh....22..323E}. Alternatively, such indexes
can also be produced in magnetic-reconnection events \citep{2013MNRAS.431..355G,2003ApJ...589..893L,2005MNRAS.358..113L,
2009MNRAS.395L..29G,2016MNRAS.462.3325P}.

During the $\gamma$F state the electron spectrum hardens, with $p\simeq 2.0$–$2.3$. This is consistent with temporarily more efficient acceleration (e.g., higher shock compression, changes in turbulence anisotropy, or a larger effective mean-free-path ratio near the shock). The modeling does not show a significant increase in $\gamma_{\rm max}$ during $\gamma$F, consistent with the relatively unchanged X-ray flux during these periods. As the other parameters do not change significantly, the $\gamma$-ray flares are more naturally attributed to changes in the injection/acceleration spectrum rather than to extended acceleration that raises the maximum energy or to changes in the emission-region size, parameters which remain statistically constant compared to the QS.

In the XF state, the inferred increase in $B$ aligns with the observed X-ray variability pattern: the X-ray flux rises while the photon index remains soft (see Figure~\ref{fig:lc}, panels c and d). Because the X-rays probe the high-energy tail of the synchrotron component, a higher $B$ shifts the synchrotron peak upward and produces a higher peak flux without requiring a hardening of the electron slope.

During OUF, the fitted parameters are broadly consistent with those in other states, but the injected slope spans a wider range, $p\simeq 1.8$–$2.7$. This spread is expected because the optical/UV band samples the rising part of the synchrotron component, where the observed slope is sensitive to the locations of the synchrotron peak and the cooling break. Small changes in $B$ or $\delta$ can move these features across the band, and modest curvature can be absorbed by the fit as changes in $p$. Moreover, if the $\gamma$-ray spectrum remains nearly unchanged while the optical/UV flux and shape vary, joint fits can accommodate a broader range of $p$.

In MWF states, the pattern indicates enhanced magnetic fields, increased power injected into electrons, and changes
in bulk motion that affect beaming, while $R$ and $p$ show no systematic change within
uncertainties. Thus, the geometry of the emitting region and the overall electron-slope shape appear roughly stable;
flares are mainly driven by plasma variations that modify the magnetic field and the acceleration conditions.

The inferred parameters for the emission region size $R$ in all the periods suggest that the radiation originates
from an extended region in the jet. This implies that the characteristic variability time,
computed as $t_{\rm var} = (1+z) R / (\delta c)$, is of several days across all states: for the average state,
with average $R \simeq 3.85 \times 10^{17}$ cm and average $\delta \simeq 21.9$, $t_{\rm var} \simeq 7.6$ days; in the XF state ($R \simeq 1.86 \times 10^{17}$ cm, $\delta \simeq 24.5$), $t_{\rm var} \simeq 3.3$ days; or in the UVOF state ($R \simeq 3.89 \times 10^{17}$ cm, $\delta \simeq 24.2$), $t_{\rm var} \simeq 7.0$ days, etc. This indicates that the flaring activity occurs over relatively large scales, without the rapid flux changes.

\subsection{Jet energetics}

We examine the jet energetics, and in particular the luminosity carried by electrons and magnetic field in different
activity states. We use the power ratio $\eta_B \equiv L_B/L_e$ to measure the magnetization of the jet. Because
$L_{e,B}\propto \pi R^{2}\Gamma^{2} c\,u'_{e,B}$, the common geometric/beaming factor cancels in the ratio, so $\eta_B$ primarily
shows the internal energy partition rather than beaming characteristics. Using the
averaged mean values, the power ratio for each independent state is: $\eta_B\simeq 0.1$
(electron-dominated by a factor $\sim 11$) in QS, $\eta_B\simeq 0.13$ ($\sim 7.9$) in MWF state, $\eta_B\simeq 0.12$ ($\sim 8.6$) in $\gamma$F state, $\eta_B\simeq 0.17$
($\sim 5.9$) in XF state and $\eta_B\simeq 0.16$ ($\sim 6.2$) in OUF. Thus, the jet is 
particle-dominated in all states ($\eta_B\approx 0.09 - 0.17$). The jet is closest to the equipartition during XF: a higher $B$ results in a larger $u'_B\propto B^{2}$ and hence $L_B$, while $L_e$ does not change  significantly. On the contrary in $\gamma$F, 
$\eta_B$ is lower than for XF which supports the
scenario that $\gamma$-ray brightening is mainly driven by particle injection properties, producing a steeper index, rather than by an
increased magnetization, or variation of other parameters.

\section{Conclusions} \label{conc}

We presented a comprehensive time-resolved multiwavelength study of the HSP
blazar \source, using extensive archival and processed data from the
\texttt{MMDC} platform. Using  observational data in the \gray, X-ray, optical,
ultraviolet and IR bands, we have investigated the source variability in these
bands and examined the origin of the emission through modeling the time-resolved
SEDs in the different emission states.

The analysis showed energy-dependent variability, with the fractional variability
amplitude peaking in the soft to medium X-ray regime ($0.7-0.8$) - while remaining moderate in the optical/UV ($0.35-0.55$) and HE \gray\ bands ($\simeq0.65$). This pattern aligns with expectations for HSP blazars, where synchrotron emission from HE electrons drives rapid variations in X-rays, contrasted by more stable contributions at lower energies.

Through Bayesian block segmentation of the \gray\ light curve, we generated
253 well-sampled time-resolved SEDs with quasi-simultaneous
data, classifying them into distinct activity states: QS, MWF, $\gamma$F,
XF, and OUF. The modeling of these SEDs within a one-zone SSC scenario shows
state-dependent parameter changes that are statistically significant
compared to QS. KS tests show that MWF episodes differ in $B$, $L_e$,
$\gamma_{\rm max}$, and $\delta$, while $R$ and $p$ are consistent with QS.
$\gamma$-ray flares are distinguished by a harder electron index $p$
(with other parameters not significantly changed), indicating that the
\gray\ brightening is due to changes in the  particle
injection/acceleration process leading to steeper electron spectra.
XF states show the broader changes in the parameters: higher $B$ and
$\gamma_{\rm max}$, smaller $R$, and higher $\delta$—consistent with
enhanced acceleration and magnetization driving the strong X-ray flares.
OUF states differ in $B$, $L_e$, $\gamma_{\rm max}$, $\delta$, and $p$,
but not in $R$, suggesting that optical/UV activity is produced by moderate
changes in magnetization, boost, and the HE cutoff without geometric
changes. During all the flaring periods, the jet
energetics is dominated by the electrons luminosity rather than the magnetic luminosity, with power ratios
$\eta_B \simeq 0.09-0.17$, which is approaching to equipartition
during X-ray-dominant activity.

The results obtained here show the complex interplay of
magnetic fields, particle acceleration, and relativistic beaming in producing
flares in blazar emission, highlighting the value of long-term, multi-epoch
modeling for these events for a better understanding of the origin of the flares.

\section*{Acknowledgements}
The research was supported by the Higher Education and Science Committee of MESCS RA (Research project No 23LCG-1C004). MK acknowledge the support by the Higher Education and Science Committee of the Republic of Armenia, in the frames of the
research project No 24AA-1C039.

We acknowledge the use of services from the Markarian Multiwavelength Data Center (\url{www.mmdc.am}).

\section*{Data Availability}
All the data used in this paper is available from \texttt{MMDC} (\url{www.mmdc.am}). The data is also available upon reasonable request to the corresponding author.



\bibliographystyle{mnras}
\bibliography{biblio.bib} 




\appendix





\bsp	
\label{lastpage}
\end{document}